\input{aipcheck}

\documentclass[
    ,final            
  ]
  {aipproc}

\layoutstyle{6x9}

\begin{document}
\url{}
\title{Vortices in self-gravitating disks}


\author{G. R. Mamatsashvili}{
  address={SUPA, Institute for Astronomy, University of Edinburgh,
Blackford Hill, Edinburgh EH9 3HJ, Scotland; E-mail: grm@roe.ac.uk},
  altaddress={Georgian
National Astrophysical Observatory, Il. Chavchavadze State
University, 2a Kazbegi Ave., Tbilisi 0160, Georgia},
email={grm@roe.ac.uk},}

\iftrue

\author{W. K. M. Rice}{
  address={SUPA, Institute for Astronomy, University of Edinburgh,
  Blackford Hill, Edinburgh EH9 3HJ, Scotland},
  email={wkmr@roe.ac.uk},}

\copyrightyear{2008}

\begin{abstract}
Vortices are believed to greatly help the formation of km sized
planetesimals by collecting dust particles in their centers.
However, vortex dynamics is commonly studied in non-self-gravitating
disks. The main goal here is to examine the effects of disk
self-gravity on the vortex dynamics via numerical simulations. In
the self-gravitating case, when quasi-steady gravitoturbulent state
is reached, vortices appear as transient structures undergoing
recurring phases of formation, growth to sizes comparable to a local
Jeans scale, and eventual shearing and destruction due to
gravitational instability. Each phase lasts over 2-3 orbital
periods. Vortices and density waves appear to be coupled implying that, in
general, one should consider both vortex and density wave modes for
a proper understanding of self-gravitating disk dynamics.

Our results imply that given such an irregular and rapidly changing,
transient character of vortex evolution in self-gravitating disks
it may be difficult for such vortices to effectively trap dust
particles in their centers that is a necessary process towards
planet formation.
\end{abstract}

\maketitle


\section{introduction}

It is well known that antcyclonic vortices can help the planet
formation process by aggregating dust particles in their centers to
build planetesimals \cite{BS95, JAB04, KB06}. Numerical simulations
\cite{GL99, UR04, JG05} demonstrate that coherent anticyclonic
vortices indeed emerge in disks and survive for hundreds of orbits.
All these investigations are, however, carried out for
non-self-gravitating disks. Here we investigate the effects of disk
self-gravity on the vortex formation and evolution, because
protoplanetary disks are in general self-gravitating and usually do
not cool fast enough to get directly fragmented into giant planets.
Instead, they settle into a quasi-steady gravitoturbulent state
\cite{Boleyetal06}. Hence, there should be found some mechanism that
will build planetesimals in this state. Global simulations of the
dynamics of dust particles in self-gravitating gaseous disks show
that large scale spiral structure in a self-regulated state does
concentrate dust particles in overdense/overpressure spiral arms
\cite{Riceetal06}. As mentioned, another possibility of dust
particle concentration is their trapping by anticyclonic vortices
(worked out originally for non-self-gravitating disks). So, in
perspective our present study will allow us to see if the latter
mechanism of planetesimal formation can also be at work in
self-gravitating disks. Due to resolution constraints, it is
difficult to see vortices in global disk simulations. For this
purpose we work in the local shearing sheet approximation.

\section{physical model and equations}

In the shearing sheet model only a local patch of a disk in the
vicinity of some radius $r_0$ is considered that rotates around the
central star with the angular velocity $\Omega_0\equiv
\Omega_K(r_0)$, where $\Omega_K(r)$ is the angular velocity of
Keplerian (differential) rotation. Within this patch the
differential rotation of a disk manifests itself as a parallel shear
flow with a constant velocity shear \cite{GL65}. The unperturbed
background surface density $\Sigma_0$ and two-dimensional pressure
$P_0$ corresponding to this shear flow are assumed to be spatially
constant. Coriolis force is also included to take into account the
effects of rotation. As a result, in this local approximation the
continuity equation and equations of motion take the form
\cite{G01}:
\begin{equation}
\frac{\partial \Sigma}{\partial t}+\nabla\cdot(\Sigma{\bf u})
-q\Omega_0 x \frac{\partial \Sigma}{\partial y} = 0,
\end{equation}
\begin{equation}
\frac{\partial {\bf u}}{\partial t}+({\bf u}\cdot\nabla){\bf
u}-q\Omega_0 x \frac{\partial {\bf u}}{\partial y} =-\frac{\nabla
P}{\Sigma}-2\Omega_0{\bf \hat{z}}\times {\bf u}+q\Omega_0u_x{\bf
\hat{y}}-\nabla \psi.
\end{equation}
This set of equations is supplemented by Poisson's equation for a
razor-thin disk
\begin{equation}
\Delta \psi=4\pi G (\Sigma-\Sigma_0)\delta(z).
\end{equation}
Here ${\bf u}(u_x, u_y), P, \Sigma$ and $\psi$ are, respectively,
the perturbed velocity relative to the background parallel shear
flow ${\bf u_0}(0, -q\Omega_0 x)$, the two-dimensional pressure, the
surface density and the gravitational potential of the gas sheet.
$x$ and $y$ are, respectively, the radial and azimuthal coordinates.
${\bf \hat{y}}$ and ${\bf \hat{z}}$ are the unit vectors in the
azimuthal and vertical directions, respectively. Since (1-2) are
written for perturbed velocities, only the gravitational potential
due to the perturbed surface density $\Sigma-\Sigma_0$ is used. The
shear parameter $q=1.5$ for the Keplerian rotation considered here.

The equation of state is
$$
P=(\gamma-1)U,
$$
where $U$ and $\gamma$ are the two-dimensional internal energy and
adiabatic index, respectively. We will adopt $\gamma=2$. The sound
speed is $c_s^2=\gamma P/\Sigma=\gamma(\gamma-1)U/\Sigma$.

The central quantity of this study is the vertical component of
potential vorticity referred to as PV below:
$$
I\equiv\frac{{\bf \hat{z}}\cdot \nabla\times {\bf
u}+(2-q)\Omega}{\Sigma}=\frac{1}{\Sigma}\left(\frac{\partial
u_y}{\partial x}- \frac{\partial u_x}{\partial y}+(2-q)\Omega
\right).
$$
The PV will play an important role in the subsequent analysis, as it
generally characterizes the formation of coherent structures
(vortices) in a disk flow \cite{JG05}.

The evolution of the internal energy density is governed by the
equation
\begin{equation}
\frac{\partial U}{\partial t}+ \nabla\cdot(U{\bf u})-q\Omega_0 x
\frac{\partial U}{\partial y} = -P\nabla\cdot{\bf
u}-\frac{U}{\tau_{c}},
\end{equation}
where the last term on the rhs takes account of cooling of the disk.
The cooling time $\tau_{c}$ is assumed to be constant,
$\tau_c=20\Omega^{-1}$, so that the disk does not fragment and
enters a saturated gravitoturbulent state. \emph{In the present
study we concentrate on examining the peculiarities of potential
vorticity evolution in such a gravitoturbulent state}.

We introduce the nondimensional variables: $t\rightarrow \Omega_0 t,
(x,y)\rightarrow (x\Omega_0/c_{s0}, y\Omega_0/c_{s0}), \Sigma
\rightarrow \Sigma/\Sigma_0, P\rightarrow P/c_{s0}^2\Sigma_0,
U\rightarrow U/c_{s0}^2\Sigma_0, I\rightarrow I\Sigma_0/\Omega_0$.
These nondimensional variables are used throughout what follows. The
Toomre's parameter is $Q=c_s\Omega_0/\pi G\Sigma$. We start with
$Q=Q_0=c_{s0}\Omega_0/\pi G\Sigma_0=1.5$.

Our computational domain in the $(x,y)$ plane is a square $-L/2\leq
x,y \leq L/2$, divided into $N\times N$ grid cells. We take $L=20$
and $N=1024$. In order to study the evolution of the system we
numerically integrate (1-4) within this domain.

\begin{figure}
\includegraphics[height=0.27\textheight]{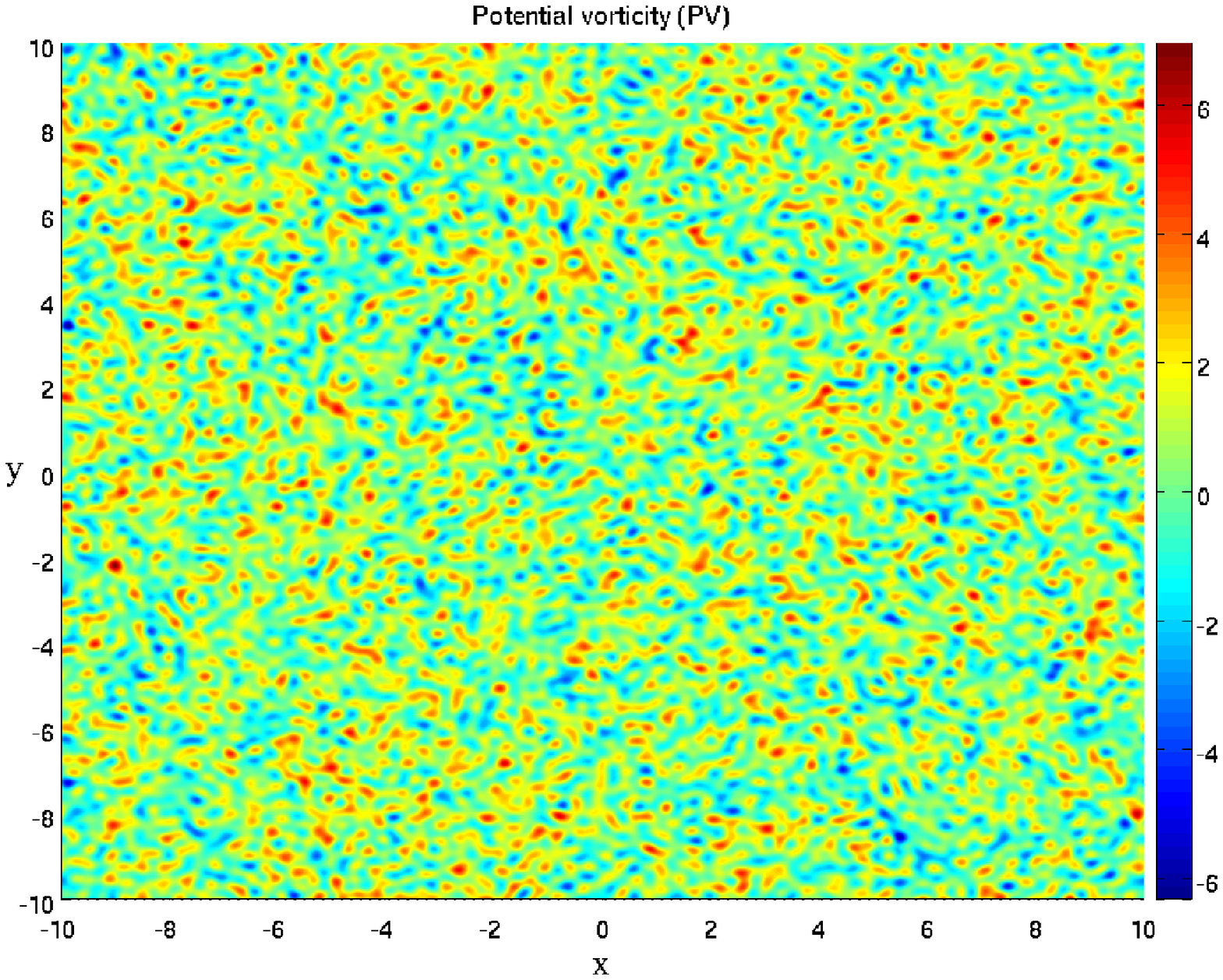}
\includegraphics[height=0.27\textheight]{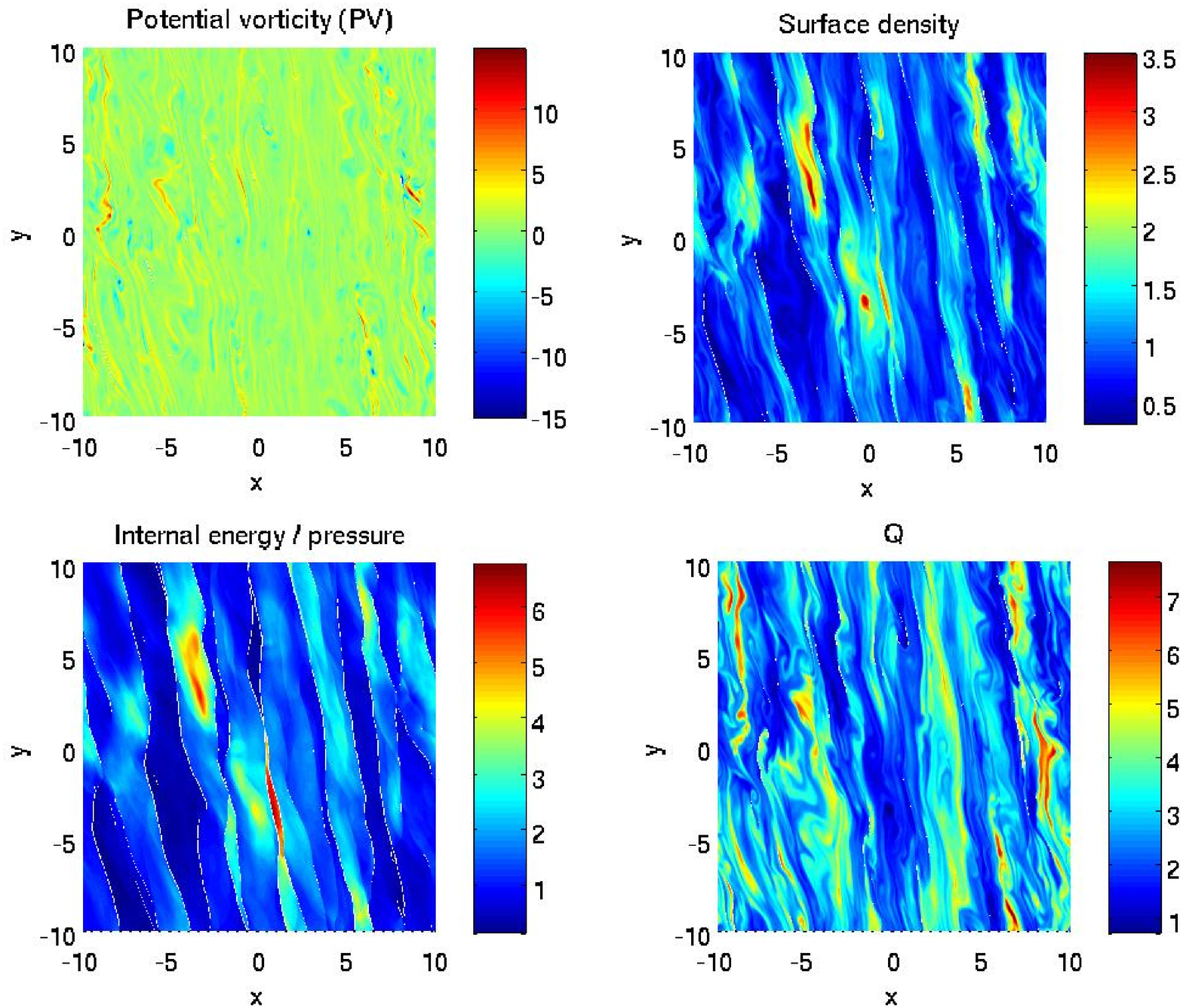}
\caption{Left: Initial PV field at $t=0$ corresponding to Kolmogorov
spectrum of velocities. Right four panels: PV, surface density,
internal energy/pressure and Toomre's parameter $Q$ in the
gravitoturbulent state at $t=33$ and $L=20$ (evolution picture
remains unchanged for larger times). Adjusted negative PV regions
produce overdense regions, which are gravitationally unstable.
Unadjusted negative PV regions correspond to central underdense
regions surrounded by overdense regions, though not so strong as for
adjusted PV regions.}
\end{figure}

\section{Nonlinear evolution}

Initial conditions consist of random $u_x$ and $u_y$ perturbations
superimposed on the mean Keplerian shear flow. Surface density and
internal energy are not perturbed initially. Fig. 1 shows these
initial conditions in terms of PV. The velocity perturbations are
measured by $\sigma = \langle {\bf u}^2(x,y)\rangle^{1/2}$, where
the angle brackets mean ensemble averaging. In our calculations
$\sigma=0.6$ at $t=0$. We start with Kolmogorov power spectrum
$\langle |u(k)|^2\rangle\sim k^{-8/3}$, where $k$ is the wavenumber.
These random velocity perturbations are meant to mimic an initial
turbulent state in a disk resulting from the collapse of a molecular
cloud core.

In the presence of both Keplerian shear and self-gravity, the main
mechanism responsible for the growth of initial velocity
perturbations is swing amplification instead of pure Jeans
instability \cite{G01, KO01, MC07}. During swing amplification
velocity perturbations induce strong surface density perturbations
in the form of trailing shocks with superimposed density structures.
After about 4-5 orbital periods balance is reached between shock and
compressional heating and cooling. As a result, the disk settles
down to a quasi-steady gravitoturbulent state. The snapshot (at
$t=33$) of the system evolution in this state is shown in fig.1. $Q$
fluctuates around $2.4$, but the $Q(x,y)$ map is very inhomogeneous
and contains values as small as $0.6$ associated with some negative
PV regions (see below). The positive (cyclonic) PV regions remain
sheared into strips showing no signs of vortex formation during the
entire course of evolution. Only negative (anticyclonic) PV regions
are able to survive in shear flows and wrap up into more or less
vortex-like structures. The overall picture of the PV evolution is
still irregular and chaotic in the quasi-steady phase (fig. 1). So,
we use the term 'vortex' in a broader sense meaning negative PV
regions in general even if they do not have well-defined vortical
shape. Some of the vortices by this time are not adjusted yet, i.e.,
they produce underdense regions corresponding to the centers of
vortices surrounded by higher density regions related to density
waves/shocks generated during the adjustment process. At the same
time, we also see in this figure vortices that have already
undergone adjustment phase, have grown to sizes comparable to the
local Jeans scale, and correspond to stronger overdense regions. At
the location of these overdense regions, $Q$ reaches small values
(0.6-0.7) implying that they are gravitationally unstable and are in
the process of being sheared and destroyed. (Vortex growth in size
is, in general, a consequence of inverse energy cascade in 2D
turbulence). During this process the temperature/internal energy
rises, the corresponding region becomes stable and the vortex
formation process described above starts again.

In conclusion, in self-gravitating disks the evolution of vortices
has irregular and transient character in contrast to that in
non-self-gravitating disks. Vortices form, undergo adjustment phase
and finally appear as overdense regions in the surface density
field, which afterwards become gravitationally unstable and are
destroyed shortly. After that the whole process recurs.


\begin{theacknowledgments}
G.R.M. would like to acknowledge the financial support from the
Scottish Universities Physics Alliance (SUPA). The numerical code
used here was kindly provided by C. Gammie.
\end{theacknowledgments}



\bibliographystyle{aipproc}   



\end{document}
\endinput